\documentclass[%
 reprint,
 amsmath,amssymb,
 aps,
 prc,
]{revtex4-2}

\usepackage{graphicx}
\usepackage{dcolumn}
\usepackage{bm}
\usepackage{hyperref}

\begin{document}

\preprint{APS/123-QED}

\title{From $\alpha$ decay to cluster decay: an extreme case of transfer learning}

\author{Yinu Zhang}
\email{zhangyn226@mail.sysu.edu.cn}
\author{Zhiyi Li}
\author{Kele Li}
\author{Jiaxuan Zhong}
\author{Cenxi Yuan}
\email{yuancx@mail.sysu.edu.cn}
\affiliation{%
 Sino-French Institute of Nuclear Engineering and Technology, Sun Yat-Sen University, Zhuhai 519082, China
}%

\date{\today}

\begin{abstract}
When training data are limited, data-driven models are especially vulnerable to optimization-related fluctuations from random initialization and to sampling-induced bias from insufficient training data. We address both challenges with transfer learning (TL): deep neural networks (DNNs) are first pretrained on $\alpha$ decay half-lives and then fine-tuned on a small cluster decay dataset. The pretraining stage provides a physically informed initialization that stabilizes optimization, while transferred global decay systematics regularize the fit and reduce sensitivity to training set composition. Despite extreme data sparsity, the resulting models accurately predict cluster decay half-lives for parent nuclei from $^{221}$Fr to $^{242}$Cm. We further quantify how initialization and sample selection affect predictive accuracy and robustness, demonstrating that TL enables stable and reliable learning in the small-sample regime.
\end{abstract}

\maketitle


\section{Introduction}

Machine learning (ML) provides broadly useful tools for scientific research~\cite{rmpML}. In nuclear physics, however, data scarcity remains a major bottleneck for ML applications. Two related issues often limit the reliability of data-driven models. First, random initialization can lead the optimization to different local minima with similar training losses but different predictive performance~\cite{jiangwg2019,losssurf}. Second, insufficient or unevenly distributed training data can introduce sampling-induced bias and weaken generalization~\cite{bias}. TL can help reduce these effects when the source and target domains share relevant underlying structure. In this approach, models learned from a data-rich source task are reused to initialize and constrain learning in a data-scarce target task, thereby improving optimization stability and reducing sensitivity to limited sample selection. The core idea of TL has been recognized in the artificial neural network community since 1976~\cite{bozinovski2020reminder}. Today, TL plays a major role in modern DNN systems~\cite{zhuang2020tl}. 

In recent years, TL has seen rapid adoption in nuclear physics, where pretrained models are efficiently adapted to new, data-scarce tasks. Applications include improving neutrino-interaction classification while requiring far fewer simulated events~\cite{chappell2022tl}, efficiently adapting a neutrino--carbon scattering generator to new nuclear targets, beam types, and interaction models to obtain more accurate neutrino-scattering event simulations with minimal additional data~\cite{bonilla2025tl}, and fine-tuning a DNN trained on electron--carbon scattering via TL to predict inclusive electron--nucleus cross sections for many other nuclei using only limited extra data~\cite{graczyk2025electron}. TL has also been applied to $\alpha$ decay half-life predictions by transferring information from a related $Q_{\alpha}$ prediction task~\cite{jyothish2025}. These successes demonstrate that TL can help us generalize to new nuclear processes when direct training data are scarce. Motivated by these advances, we explicitly test TL under conditions of extreme data scarcity by training a DNN on $\alpha$ decay data and adapting it to predict cluster decay half-lives, for which experimental data are exceptionally limited. This application of TL in physics illustrates its potential to improve predictive modeling in data-limited nuclear physics problems.
Cluster decay is an exotic decay mode that occurs as an extremely rare branch relative to $\alpha$ decay from the same parent nucleus. The branching ratios relative to $\alpha$ decay range from $10^{-9}$ to $10^{-16}$. Only a few types of emitted light clusters heavier than an $\alpha$ particle have been observed. The first confirmed cluster decay, the emission of a $^{14}\text{C}$ from $^{223}\text{Ra}$, was theoretically predicted in 1980~\cite{sandulescu1980} and experimentally observed in 1984~\cite{rose1984}. Subsequent studies revealed that actinides from $^{221}\text{Fr}$ to $^{242}\text{Cm}$ emit light fragments ranging from $^{14}\text{C}$ to $^{34}\text{Si}$~\cite{poenaru2002,poenaru2008}. For nuclei heavier than $^{242}\text{Cm}$, spontaneous fission becomes a dominant competing process, complicating the detection of cluster decay products among abundant fission fragments. The extreme scarcity of reliable cluster decay data makes robust direct training difficult, which is a prime motivation for adopting a TL approach.

Physically, $\alpha$ decay and cluster decay are closely related. Both processes can be described as charged-particle tunneling through the Coulomb barrier; cluster decay can therefore be viewed as a rare extension of $\alpha$ decay, governed by similar mechanisms but with heavier emitted fragments and larger $Q$ values. Several theoretical models have unified these two decay modes. Qi \textit{et al}.~\cite{qi2009udl,qi2009micro} derived a universal decay law (UDL) within an $R$-matrix framework that generalizes the Geiger--Nuttall law from $\alpha$ decay to heavier emitted clusters. The UDL establishes a linear relationship between the logarithm of the half-life and combinations of the $Q$ value, the masses and charges of the emitted particles, and those of the parent nuclei. It was shown to describe all known cluster decays and $\alpha$ decays within a single framework. Similarly, Poenaru \textit{et al}.~\cite{poenaru2011univ} obtained a single universal line for $\alpha$ decay and cluster decay by plotting the sum of the decimal logarithm of the half-life and the cluster preformation probability against the decimal logarithm of the external-barrier penetrability in a fission-like framework. Ni \textit{et al}.~\cite{ni2008} proposed a unified empirical formula, derived from the WKB approximation, for calculating the half-lives of $\alpha$ decay and cluster decay. Microscopic studies have also examined cluster formation in both decay modes~\cite{delion1994,uzawa2022,zhao2023,cai2026}. This commonality provides the physical basis for our TL architecture. A model pretrained on $\alpha$ decay data can learn the key physics and then be transferred to predict cluster decay half-lives using only a small amount of cluster decay data for fine-tuning.

ML methods have been widely applied to $\alpha$ decay half-life prediction in recent years~\cite{saxena2021,panglonggang2022,zhanghongfei2022,cai2023random,liujian2023,manana2023,cheng2024alpha,jalili2024,hexiaotao2024,shree2025alpha,manana2025,zhanghaifei2025,jyothish2025,hexiaotao2025}. These data-driven models have shown strong predictive capability across the nuclear landscape, and several studies report that their predictive performance is comparable to traditional nuclear physics models. These successes in $\alpha$ decay modeling provide a solid foundation for applying TL to the data-scarce problem of cluster decay.

The rest of this paper is organized as follows. Section~\ref{methodology} describes the methodology, including the dataset, input features, neural-network architecture, and the TL fine-tuning procedure from $\alpha$ decay to cluster decay. Section~\ref{results} presents a detailed discussion of how TL improves predictive accuracy. Section~\ref{summary} summarizes our conclusions and outlines directions for future research.

\section{Methodology\label{methodology}}

This study aims to explore the potential of TL for training a DNN to predict cluster decay half-lives. Our approach first pretrains a DNN in the $\alpha$ decay domain and then fine-tunes it in the cluster decay domain with very few training examples. This strategy transfers knowledge from the $\alpha$ decay channel to the much rarer cluster decay regime, thereby reducing overfitting, mitigating optimization-related fluctuations and sampling-induced bias, and improving generalization under severe data scarcity.

\begin{figure*}
 \includegraphics[width=\textwidth]{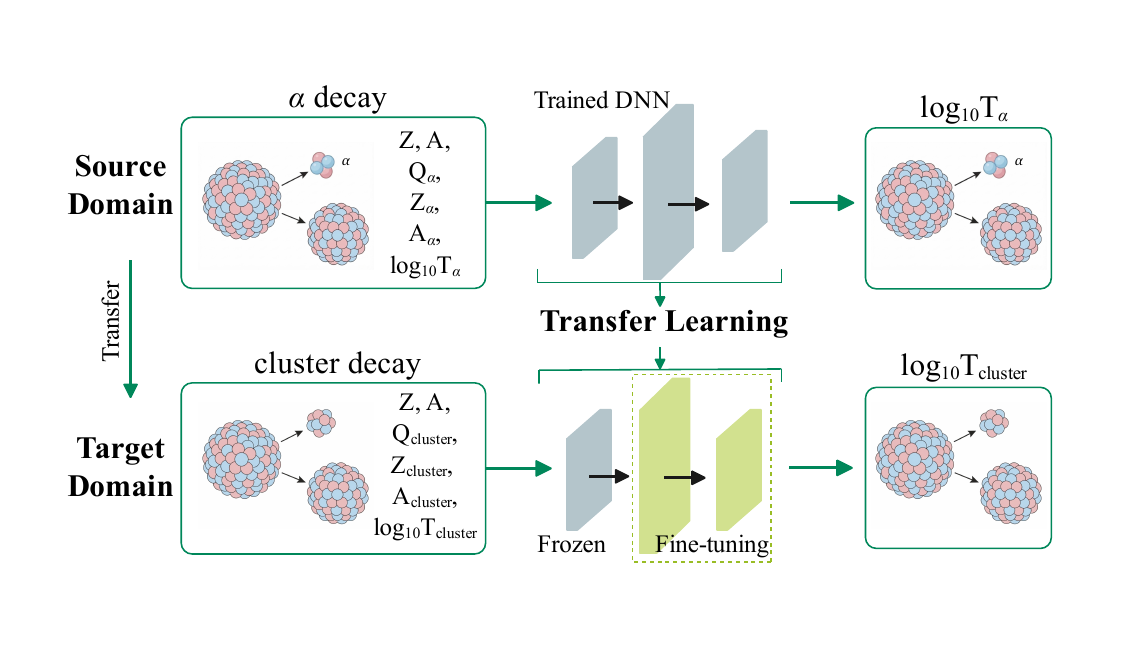}\\
 \caption{Schematic of the TL architecture. In general, TL reuses knowledge learned from a data-rich source task to improve performance on a related, data-scarce target task. In this work, a DNN is first pretrained on the abundant $\alpha$ decay dataset and then adapted to cluster decay by either fine-tuning all layers (full fine-tuning) or fine-tuning only the last few layers (shallow fine-tuning), thereby leveraging shared tunneling systematics while limiting overfitting in the data-scarce target domain.}
 \label{architecture}
\end{figure*}

\subsection{Pretraining}

To enable TL from $\alpha$ decay to the far more data-scarce cluster decay, we first pretrain a DNN to regress experimental logarithmic $\alpha$ decay half-lives, $y=\log_{10}T_{1/2}$, following Ref.~\cite{manana2023}. The input feature set is defined as $\boldsymbol{x}=\{Z_d, A_d, Z_c, A_c, Q\}$, where $Z_d$ and $A_d$ are the charge and mass numbers of the daughter nucleus, $Z_c$ and $A_c$ are the charge and mass numbers of the emitted $\alpha$ particle or cluster, and $Q$ is the corresponding decay $Q$ value.

The DNN, denoted $f(\mathbf{x};\boldsymbol{\theta})$, comprises an input layer and one or more fully connected hidden layers and an output layer. For a hidden layer with tanh activation, the output is written as
\begin{equation}
y=a+\sum_{j=1}^{H}b_{j}\tanh\!\left(c_j+\sum_{i=1}^{N}\omega_{ji}x_i\right),
\end{equation}
where $\boldsymbol{\theta}\equiv\{a,b_j,c_j,\omega\}$ collects all weights and biases. These free parameters critically determine the DNN’s predictive performance and generalization. Random initialization often causes different runs to follow distinct optimization trajectories and converge to different local minima, resulting in fluctuations and inconsistent generalization performance. $N$ is the number of feature dimensions. Both the hidden-layer width $H$ (the number of neurons per hidden layer) and the network depth (the number of hidden layers) are treated as hyperparameters. During pretraining on $\alpha$ decay data, $\boldsymbol{\theta}$ is optimized via backpropagation using the Levenberg--Marquardt algorithm; the same optimizer is used during TL~\cite{more2006LM,zhanghongfei2017,manana2023}. The Levenberg--Marquardt algorithm is a standard and efficient second-order type optimization method. It is particularly suitable for the present problem because the network contains only a limited number of parameters, and for the present task, it performs better than Adam~\cite{kingma2015adam} and SGD~\cite{bottou2010sgd}.
\begin{equation}
\boldsymbol{\theta_{i+1}} = \boldsymbol{\theta_{i}}-(J^T J + \lambda I)^{-1} J^T r,
\end{equation}
Here, $J$ is the Jacobian matrix of partial derivatives of the residuals with respect to the parameters, $I$ is the identity matrix, and $r$ is the residual vector defined by the difference between predicted and experimental values. The adaptive tuning parameter $\lambda$, which plays a role analogous to the learning rate, controls the damping of the Levenberg--Marquardt updates. During pretraining, smaller $\lambda$ values allow larger curvature-guided updates. By contrast, during the TL stage, using a larger $\lambda$ yields a more conservative damping strategy that restricts the magnitude of parameter updates, thereby mitigating overfitting and preventing catastrophic forgetting of pretrained knowledge during adaptation to the target task. The model minimizes the least squares objective, $D$, in two stages: first, through pretraining on $\alpha$ decay systematics, which provides a physically informed initialization that regulates the minimization process; and second, through TL by minimizing $D$ on cluster decay training data.
\begin{equation}
D=\sum_{m=1}^{N_s}(y_m-t_m)^2,
\end{equation}
with $t_m=\log_{10}T_{1/2,m}^{\rm Exp.}$, where $N_s$ denotes the number of samples in the dataset. Predictive accuracy is quantified by the root-mean-square (rms) deviation
\begin{equation}
\sigma_{\mathrm{rms}}
=\left\{\frac{1}{N_s}\sum_{i=1}^{N_s}\left[\log_{10}\!\left(\frac{T_i^{\mathrm{Cal.}}}{T_i^{\mathrm{Exp.}}}\right)\right]^2\right\}^{1/2}.
\end{equation}

For pretraining, we use 591 ground-state $\alpha$ decay half-lives from Ref.~\cite{NNDC}, including 137 even--even, 132 even-odd, 163 odd-even, and 159 odd-odd nuclei, spanning $105\leqslant A\leqslant 294$ and $52\leqslant Z\leqslant 118$. For the present study, as in many recent ML applications for decay studies, the available $\alpha$ decay dataset remains limited for training modern DNNs. To keep the model complexity in line with the available data, we therefore use compact neural network architectures with only a small number of hidden neurons, rather than large overparameterized DNNs. This choice reduces the risk of overfitting and makes the present analysis appropriate for the scale of the ground-state $\alpha$ decay dataset. In addition, the present work focuses only on ground-state to ground-state cluster decay. For consistency, the source domain data used for pretraining are likewise restricted to ground-state to ground-state $\alpha$ decay. This ensures that the transferred information is based on comparable decay observables and avoids introducing additional effects from excited final states, angular momentum barriers, spin-parity changes, and branching ratios. The present work should therefore be regarded as a proof-of-principle study of transfer learning under extreme target domain data scarcity. To control overfitting, we first randomly split the $\alpha$-decay data into an 80\% training set and a 20\% held-out test set. We then perform 10-fold cross-validation only within the 80\% training set. Specifically, the training set is divided into 10 disjoint folds, and the model is trained 10 times, each time using 9 folds for parameter optimization and the remaining fold for validation. The held-out 20\% test set is excluded from cross-validation and reserved for the final evaluation of the selected model. The mean $\sigma_{\rm rms}$ across folds is used to evaluate the pretrained model performance.  We explored a large hyperparameter space over different network depths and widths, and by jointly considering pretraining, full fine-tuning, and shallow fine-tuning performance, we found that compact architectures provide the best balance between source task fitting capacity and target task generalization. In particular, although larger networks can improve the fitting performance in the $\alpha$ decay pretraining task, they are less suitable for subsequent transfer to the extremely small cluster decay dataset. We therefore selected the final architectures by jointly considering the pretraining performance on $\alpha$ decay and the transfer performance on cluster decay, reducing the risk of overfitting to the training set. The resulting choices are: (i) a network with one hidden layer of six neurons for full fine-tuning, and (ii) a network with two hidden layers of six neurons each for shallow fine-tuning. The $\alpha$ decay test set results are used for architecture selection and consistency checks. Once the architecture is fixed, the DNN is retrained on the full $\alpha$ decay dataset to obtain the pretrained parameter set $\boldsymbol{\theta}_{\rm pre}$ used for subsequent transfer.

The pretrained parameter set $\boldsymbol{\theta}_{\rm pre}$ is then used to initialize the network for the target task, replacing a random initialization with a physically informed starting point learned from the much larger $\alpha$ decay dataset. Since $\boldsymbol{\theta}_{\rm pre}$ already captures the shared underlying physics of $\alpha$ decay and cluster decay, quantum tunneling of a charged particle through the Coulomb barrier, this initialization transfers that knowledge from the data-rich $\alpha$ decay domain to the cluster decay task. As a result, it enhances optimization stability and improves generalization when only limited cluster decay data are available.
\begin{figure*}
 \includegraphics[width=\textwidth]{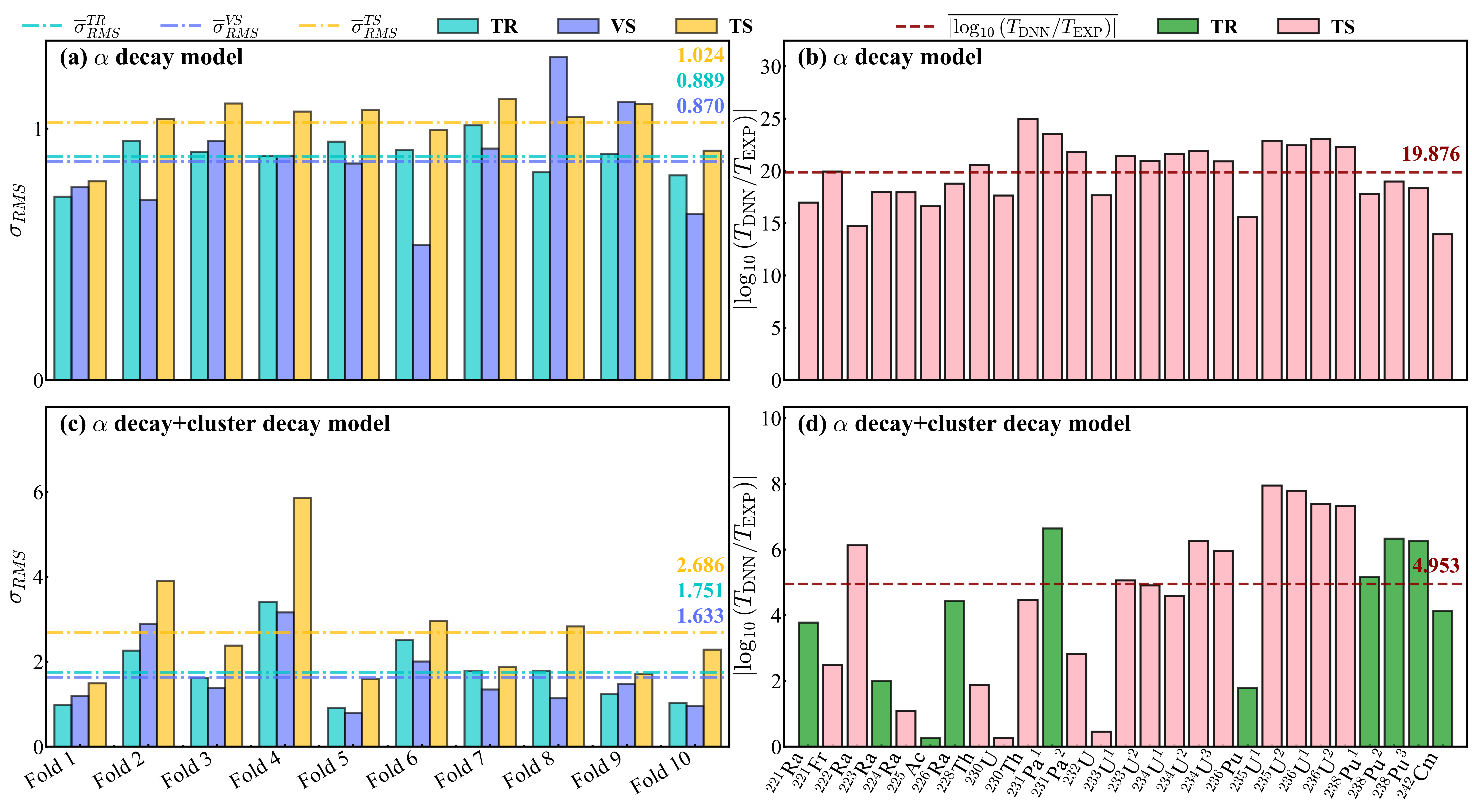}\\
 \caption{Performance evaluation using 10-fold cross-validation. The left panels show the rms deviations, $\bar{\sigma}_{\rm rms}$, for the training (TR), validation (VS), and test (TS) folds in the $\alpha$ decay (a), and $\alpha$+cluster decay (c). The right panels illustrate the domain shift when an $\alpha$-only network is applied directly to cluster decay (b) and the corresponding performance when the network is trained on the combined $\alpha$+cluster decay dataset (d). The x-axis of the right panels denotes the parent nuclei for cluster decay, where the right superscripts demonstrate different decay channels, green strips are training data, and pink strips are test data.}
 \label{dnn_evaluation}
\end{figure*}

\subsection{Transfer Learning}

After pretraining on the $\alpha$ decay dataset, we fine-tuned the model using cluster decay data. The architecture of this study is illustrated in Fig.~\ref{architecture}. Inspired by the human ability to transfer knowledge across domains, TL seeks to leverage information from a related source domain to improve performance and reduce the number of labeled examples required in a target domain. However, transferred knowledge is not always beneficial. If the source and target domains share little common structure, transfer can fail or even degrade performance~\cite{zhuang2020tl}. The target task here is the prediction of heavy-cluster emission half-lives, for which we use 27 experimentally confirmed data points from Ref.~\cite{bonetti2007cluster}. To systematically evaluate the efficacy of TL for cluster decay prediction, we investigate the transfer process through two distinct fine-tuning regimes:

1. Full Fine-Tuning:
Starting from the pretrained $\alpha$ decay model, all weights and biases in the network are allowed to be re-optimized using the cluster decay training data.

2. Shallow Fine-Tuning:
We impose a constraint that freezes the weights of the initial and intermediate layers at their pretrained values from the $\alpha$ decay model, and restrict backpropagation updates exclusively to the parameters of the last few layers. In this work, the parameters of the last 2 layers (the final hidden layer and the output layer) are re-optimized.

After fine-tuning, we obtained a DNN specialized for predicting cluster decay half-lives while retaining the generalization capability and optimization stability learned from $\alpha$ decay. The model carries into the cluster decay domain an internal physics representation learned from $\alpha$ decay. The transfer is crucial because the limited number of available cluster decay data points makes direct training from random initialization highly unstable and weakly generalizing. TL therefore provides a physically informed starting point and a global regulator, enabling reasonable predictions from the outset and allowing subsequent refinement with only minimal parameter updates.

\section{Results and discussion\label{results}}
First, we use $k$-fold cross-validation to evaluate the generalization performance of DNNs trained on two datasets: one containing only the $\alpha$ decays training set, and the other combining both $\alpha$ and cluster decays training set. We then assess the trained models' ability to predict cluster decay half-lives. $k$-fold cross-validation improves data efficiency because each nucleus is used once for validation and in the remaining folds for training, which is particularly important in nuclear physics applications with limited experimental data. Moreover, averaging validation errors across folds reduces sensitivity to any single split and yields a more robust estimate of predictive accuracy. In Fig.~\ref{dnn_evaluation}(a), the $\alpha$ decay training set is randomly divided into 10 equal folds; the model is trained on 9 folds and validated on the remaining one. The resulting DNN achieves high training and validation accuracy, as well as good generalization on the $\alpha$ decay test set. In Fig.~\ref{dnn_evaluation}(b), we apply the $\alpha$ only DNN directly to cluster decay; the model fails catastrophically because the network is trained only on $\alpha$ decays, where emitted-particle features are fixed at $Z_c=2$ and $A_c=4$, whereas cluster decay involves much larger values of $Z_c$, $A_c$, and $Q$. Nevertheless, this failure in direct application does not imply that pretraining is ineffective; the learned $\alpha$ decay systematics still provide a physically informed initialization for transfer learning, which improves optimization stability and generalization after fine-tuning on cluster decay data. In Fig.~\ref{dnn_evaluation}(c), we add 10 cluster decay half-lives to the training set, assign the remaining data to the test set, then perform the 10-fold cross-validation to evaluate generalization on the combined training set. In this combined model, the much larger $\alpha$ decay dataset overwhelms the tiny cluster decay dataset. The combined model, therefore, primarily learns $\alpha$ decay systematics; moreover, adding a few cluster decay samples degrades performance on $\alpha$ decay data and fails to correct model behavior for cluster decay. The $\sigma_{\rm rms}$ values for TR are relatively small but increase for VS and TS. Moreover, although $\alpha$ decay and cluster decay share the same underlying physics, their feature ranges differ substantially because cluster decay has much larger $Z_c$, $A_c$, and $Q$ values; optimization is therefore strongly shaped by the $\alpha$ decay data. As shown in Fig.~\ref{dnn_evaluation}(d), the combined model performs poorly for cluster decay. In contrast, TL uses the $\alpha$ decay data only for pretraining. During fine-tuning, the loss function is evaluated only on the cluster decay training data, so the target domain data are not overwhelmed by the source domain data but instead adapt the pretrained model to cluster decay.

\begin{figure}
 \includegraphics[width=\columnwidth]{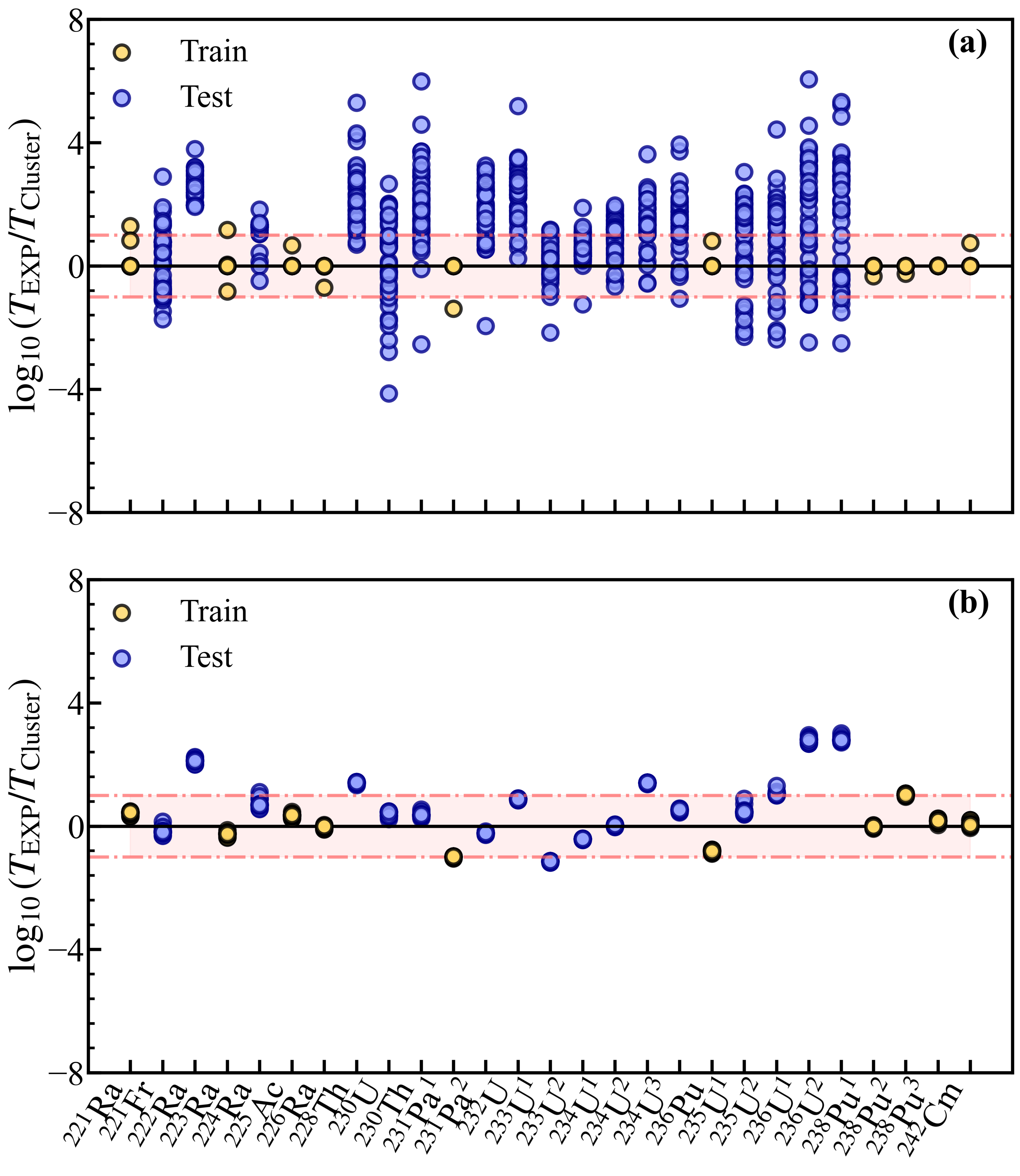}\\
 \caption{Optimization-related fluctuations for half-life prediction are shown using the same cluster decay train/test data in Fig.~\ref{dnn_evaluation}. The x-axis denotes the parent nuclei for cluster decay, where the right superscripts demonstrate different decay channels, and the y-axis denotes the difference between the experimental value and the corresponding TL prediction. Panel (a) shows a model trained directly on 10 cluster decay samples from random initialization, which overfits and exhibits large variability across 50 random initializations. Panel (b) shows the full TL model initialized with pretrained $\alpha$ decay parameters $\boldsymbol{\theta}_{\rm pre}$ and then fine-tuned on the cluster decay training subset, yielding more stable predictions for both training and test data. Panel (c) compares the standard deviations of the residuals over the 50 runs for direct training and TL, quantifying the suppression of optimization-related fluctuations by transfer learning.}
 \label{optimization_fluctuations}
\end{figure}

TL provides a physically informed pretrained parameter set $\boldsymbol{\theta}_{\rm pre}$ to mitigate optimization-related fluctuations caused by random initialization. Fig.~\ref{optimization_fluctuations}(a) and Fig.~\ref{optimization_fluctuations}(b) use the same compact network with only one hidden layer of six neurons, and the cluster decay training and test splits are identical to those used in Fig.~\ref{dnn_evaluation}. In Fig.~\ref{optimization_fluctuations}(a), the model is trained directly on only 10 cluster decay samples from random initialization, overfits the training set, and shows widely varying generalization. This behavior is consistent with the nonconvex nature of DNN optimization, where many parameter configurations can achieve similarly low training loss~\cite{losssurf}. With a small target dataset, the empirical objective is weakly constrained, so different random initializations can converge to distinct low-loss basins with markedly different test errors. Consequently, multiple solutions may fit the training data comparably well yet generalize very differently, as shown in Fig.~\ref{optimization_fluctuations}(a), which makes robust direct training difficult. With larger and more representative datasets, initialization-induced variability is typically reduced because the objective constrains the solution more strongly. TL turns this underconstrained learning problem into a better constrained adaptation problem. By introducing a strong inductive bias toward generalizable features learned from the $\alpha$ decay dataset, TL reduces the set of low-loss solutions; in Fig.~\ref{optimization_fluctuations}(b), the 50 runs no longer start from arbitrary random weights; instead, the initialization is constrained by $\alpha$ decay pretraining, which provides the physically informed pretrained parameter set $\boldsymbol{\theta}_{\rm pre}$. This constrained initialization leads the full fine-tuning process to converge toward similar, well-generalizing solutions. Consequently, test performance is stable and reliable, yielding $\sigma_{\mathrm{rms}}=$ 1.089, as shown in Fig.~\ref{optimization_fluctuations}(b). To quantify optimization-related fluctuations, we fix the cluster decay training/test split and repeat the training procedure 50 times using different random initializations, as shown in Fig.~\ref{optimization_fluctuations}(c). For each nucleus $i$, let $\Delta_i^{(k)}=\log_{10} T_{1/2,i}^{\rm exp}-\log_{10} T_{1/2,i}^{\rm pred,(k)}$ denote the residual in run $k$, and let $\bar{\Delta}_i=(1/50)\sum_{k=1}^{50}\Delta_i^{(k)}$ be the mean residual over the 50 runs. We define the corresponding standard deviation as $\mathrm{SD}_i=\left[(1/50)\sum_{k=1}^{50}\left(\Delta_i^{(k)}-\bar{\Delta}_i\right)^2\right]^{1/2}$. A detailed comparison of the SD values can be seen in Fig.~\ref{optimization_fluctuations}(c). Comparing the SD values from direct training and TL quantifies how strongly TL suppresses optimization-related fluctuations. Smaller SD values indicate weaker sensitivity to random initialization and hence greater robustness. The results show that TL is much less sensitive to random initialization than direct training.

\begin{figure}
 \includegraphics[width=\linewidth]{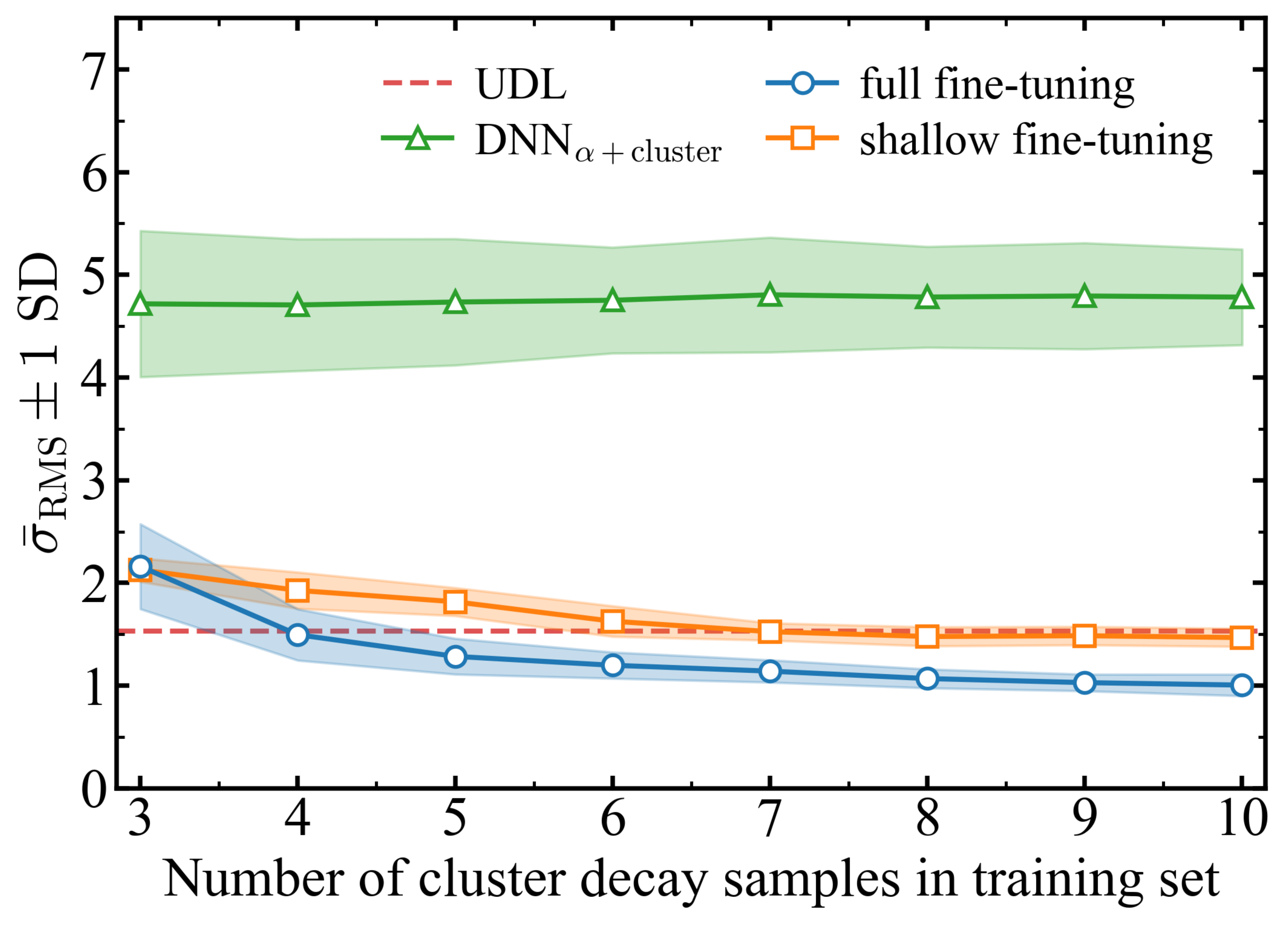}\\
 \caption{Predictive performance as a function of the number of cluster decay training samples. For each sample size, the mean $\sigma_{rms}$ and its standard deviation for complete cluster decay data were computed from 50 random training set selections, evaluating three approaches: the combined $\alpha$+cluster model, shallow TL fine-tuning, and full TL fine-tuning. The standard deviation is taken over these selections and quantifies the sensitivity to the specific nuclei are included in the small cluster decay training set. The UDL result is included as a benchmark.}
 \label{comparison}
\end{figure}

TL provides a global regulator learned from $\alpha$ decay data that mitigates sampling-induced bias from insufficient training data. To systematically compare predictive performance and assess the limitations of TL, we examine three approaches: the combined $\alpha$+cluster model, shallow TL fine-tuning, and full TL fine-tuning. We exclude direct training on the cluster decay data alone because, in the extremely small-sample regime considered here, it is highly unstable and strongly sensitive to random initialization, as shown in Fig.~\ref{optimization_fluctuations}(a). For each approach, we increase the number of cluster decay training samples from 3 to 10. At each sample size, we randomly construct 50 distinct cluster decay training subsets and evaluate the model on all 50 subsets.  This design allows us to quantify how TL performance depends on both training set size and the specific sample composition. This approach is complementary to the $k$-fold cross-validation used above: $k$-fold cross-validation estimates average generalization for a fixed dataset by rotating validation folds, whereas the analysis using 50 random combinations directly quantifies sensitivity to which specific nuclei are selected for training at each sample size. Because very small training sets can make DNN performance highly sensitive to sample selection, Fig.~\ref{comparison} reports the mean $\sigma_{rms}$ for all cluster decay data and its standard deviation across those 50 selections, with the UDL included as a benchmark. For very small training sets, the combined $\alpha$+cluster model shows a relatively modest standard deviation because abundant $\alpha$ decay data strongly constrain the network; however, its mean $\sigma_{rms}$ remains poor due to data imbalance, as the much larger number of $\alpha$ decay samples dominates optimization. Both shallow and full fine-tuning perform markedly better, with substantially smaller standard deviations across all training set sizes, indicating reduced sensitivity to sample selection and more robust generalization. Crucially, the fully fine-tuned model reaches UDL-level accuracy with only four training samples, whereas the shallow fine-tuned model requires seven. Initialization from $\boldsymbol{\theta}_{\rm pre}$ imposes a strong physics-informed inductive bias that constrains fine-tuning to a physically plausible function class. To quantify sampling-induced bias, for each training set size let $\sigma_{rms}^{(m)}$ denote the value obtained from the $m$th randomly constructed training subset, and let $\bar{\sigma}_{rms}=(1/50)\sum_{m=1}^{50}\sigma_{rms}^{(m)}$ be the mean over the 50 subsets, as summarized in Fig.~\ref{comparison}. We then define the corresponding standard deviation as $\mathrm{SD}(\sigma_{rms})=\left[(1/50)\sum_{m=1}^{50}(\sigma_{rms}^{(m)}-\bar{\sigma}_{rms})^2\right]^{1/2}$. This quantity measures how strongly the predictive performance depends on which nuclei are selected for training. Thus, for each fixed number of cluster decay training samples, the SD values in Fig.~\ref{comparison} and Table~\ref{tab:comparison_sd} quantify the variation of $\sigma_{rms}$ over 50 random choices of the target-domain training subset. This quantity should be distinguished from the initialization-induced optimization fluctuations discussed in Fig.~\ref{optimization_fluctuations}, where the data split is fixed, and only the random initialization is varied. Here, the variation reflects sensitivity to the composition of the small cluster decay training set. This sensitivity arises from the sparsity and heterogeneity of the cluster decay data: each selected decay channel carries information about a specific parent nucleus, emitted particle, and $Q$ value. When only a few cluster decay examples are available for fine-tuning, adding or removing one informative nucleus can change the target-domain correction learned from the pretrained $\alpha$ decay systematics. The reduced SD values after TL therefore indicate that pretraining on $\alpha$ decay systematics constrains the cluster decay fit, making the predictions less dependent on the specific samples used for fine-tuning. The corresponding SD values for the three models at each training set size are summarized in Table~\ref{tab:comparison_sd}.

\begin{table}
\caption{Standard deviations of $\sigma_{rms}$ for the combined $\alpha$+cluster model, full fine-tuning, and shallow fine-tuning shown in Fig.~\ref{comparison}, as a function of the number of cluster decay training samples. The SD is computed over the 50 random training set selections and measures sample selection sensitivity rather than optimization-related fluctuations from random initialization.}
\label{tab:comparison_sd}
\begin{ruledtabular}
\small
\begin{tabular*}{\columnwidth}{@{\extracolsep{\fill}}lcccccccc}
Model & 3 & 4 & 5 & 6 & 7 & 8 & 9 & 10 \\
 \hline
$\alpha$+cluster & 0.71 & 0.64 & 0.62 & 0.52 & 0.56 & 0.49 & 0.52 & 0.47 \\
Full FT & 0.41 & 0.25 & 0.17 & 0.13 & 0.11 & 0.09 & 0.08 & 0.10 \\
Shallow FT & 0.11 & 0.18 & 0.14 & 0.15 & 0.08 & 0.09 & 0.09 & 0.09 \\
\end{tabular*}
\end{ruledtabular}
\end{table}

The data efficiency of TL can be understood by mapping the network's operation to the physics encoded in the UDL~\cite{qi2009udl}. In the UDL, the half-life for charged-particle emission is written as
\begin{equation}
\log T_{1/2} = a \chi' + b \rho' + c,
\label{eq:UDL}
\end{equation}
where $\chi' = Z_c Z_d \sqrt{A/Q}$ captures the Coulomb barrier penetrability and $\rho' = \sqrt{A Z_c Z_d (A_d^{1/3} + A_c^{1/3})}$ encodes the dependence on the cluster preformation amplitude through the approximately linear behavior of $\log|R F_c(R)|$ with $\rho'$~\cite{qi2019}. For $\alpha$ decay, the formation amplitude varies little, so the half-life is dominated by the $\chi'$ term, i.e., by the Geiger--Nuttall law. For heavier clusters, by contrast, the formation probability changes by up to 8 orders of magnitude, and the UDL absorbs this large variation into the $\rho'$ term. The pretrained model for $\alpha$ decay faces a similar situation: it learns the dominant $\chi'$ dependence while treating the formation contribution as nearly constant. When the network is subsequently fine-tuned on cluster decay data, the universal barrier penetration physics encoded in the weights is largely retained, while the parameters governing the $\rho'$ pathway are adjusted globally to reproduce the formation slope appropriate for heavier clusters. In other words, fine-tuning does not need to relearn the tunneling systematics from scratch; instead, it recalibrates the probability of cluster formation. To make this transfer of physically meaningful feature representations more explicit, we also analyze both the pretrained $\alpha$ decay model and the final full fine-tuning model using SHapley Additive exPlanations (SHAP)~\cite{mumpower2022}. For each input feature $\boldsymbol{x}=\{Z_d, A_d, Z_c, A_c, Q\}$, the SHAP value measures its contribution to the predicted $\log_{10}T_{1/2}$ relative to a background expectation. By comparing the mean absolute SHAP value before and after fine-tuning, we can identify which feature contributions are inherited from pretraining and which are transferred to describe cluster decay systematics. The change in SHAP values is physically meaningful. In the pretrained model, the SHAP values of $Z_c$ and $A_c$ are zero because, in the source-domain $\alpha$ decay data, the emitted particle is always an $\alpha$ particle with fixed $Z_c=2$ and $A_c=4$, so these two inputs carry no discriminating information. After transfer to cluster decay, however, $Z_c$ and $A_c$ become much more important because the emitted clusters span a range of charges and masses. This increase is consistent with the UDL interpretation discussed above: $Z_c$ and $A_c$ enter the variable $\rho'$, which encodes the cluster formation contribution to the half-life. Their enhanced SHAP values therefore indicate that full fine-tuning captured the cluster formation contribution, which is only weakly constrained in the pretrained $\alpha$-only regime. In this sense, SHAP provides an interpretable view of feature transfer: the pretrained model supplies the dominant global decay descriptors, while full fine-tuning selectively recalibrates their relative importance in the target domain. The corresponding SHAP values for the pretrained and final full fine-tuning models are summarized in Table~\ref{tab:shap}. This interpretation also helps explain why full fine-tuning outperforms shallow fine-tuning: the formation amplitude depends nontrivially on all input features through $\rho'$, capturing it requires adjusting weights throughout the network. The fact that only four cluster decay examples are sufficient to reach UDL-level accuracy suggests that the pretrained model suggests physically meaningful directions analogous to $(\chi', \rho')$. TL thus efficiently preserves the universal penetrability information while learning the cluster-specific formation linearity. Thus, the DNN's internal representation, first shaped by $\alpha$ decay systematics and then adapted to cluster decay, aligns naturally with the physical separation between Coulomb barrier penetration and preformation probability encoded in the UDL.

\begin{table}
\caption{Mean absolute SHAP values for the five input features in the pretrained $\alpha$ decay model and the final full fine-tuning model. Comparing the two rows highlights how feature importance is transferred and recalibrated during adaptation to cluster decay.}
\label{tab:shap}
\begin{ruledtabular}
\small
\begin{tabular*}{\columnwidth}{@{\extracolsep{\fill}}lccccc}
Model & $Z_d$ & $A_d$ & $Z_c$ & $A_c$ & $Q$ \\
\hline
Pretrained & 0.065 & 0.325 & 0.0 & 0.0 & 0.610 \\
Full FT & 0.052 & 0.017 & 0.324 & 0.154 & 0.453 \\
\end{tabular*}
\end{ruledtabular}
\end{table}

\section{Summary and outlook\label{summary}}
In summary, we developed and tested a TL strategy to address extreme data scarcity in cluster decay studies. The model is first pretrained on 591 ground-state $\alpha$ decay half-lives and then fine-tuned on the limited available cluster decay data. This transfer serves two key roles: it provides a physically informed initialization that suppresses optimization-related fluctuations and a global regulator that reduces sampling-induced bias. As a result, the fine-tuned DNN achieves accurate and stable predictions of cluster decay half-lives. In particular, full fine-tuning reaches UDL-level accuracy with only four training samples, demonstrating that knowledge learned from $\alpha$ decay can be effectively reused for a much rarer decay channel. Future work will incorporate Bayesian neural networks into the present TL framework to improve uncertainty quantification. Another important future direction is to extend the framework to include both the fine structure of $\alpha$ decay and cluster decay to excited states, which would require additional descriptors such as excitation energies, spin-parity assignments, angular-momentum transfer, and branching information.

More broadly, the transfer from $\alpha$ decay to cluster decay offers a proof of concept for data-limited problems across nuclear and particle physics. The same framework can be extended to other pairs of abundant and rare processes, such as neutron capture near stability versus along the $r$-process path~\cite{liddick2016}, single $\beta$ decay versus double $\beta$ decay~\cite{suhonen2017}, and actinide fission systematics versus fission properties far from stability~\cite{giuliani2018}. Future work will focus on expanding high-quality target datasets, improving uncertainty quantification, and incorporating stronger physics constraints during fine-tuning to further enhance the reliability of extrapolation.

\section{Data availability\label{Data}}
The TL model and corresponding data are available at the GitHub repository~\cite{lizhiyi2026}.


\begin{acknowledgments}
This work was supported by the National Natural Science Foundation of China under Contracts No. 12305129, No. 12475129, Guangdong Major Project of Basic and Applied Basic Research under Grant No. 2021B0301030006, and the computational resources from Sun Yat-Sen University and Fudan University. Data associated with this work can be obtained by request to the authors.
\end{acknowledgments}

\bibliographystyle{apsrev4-2} 
\bibliography{ref}

\end{document}